\documentclass{aims}
 \usepackage{amsmath,bm}

  \usepackage{paralist}
  \usepackage{graphics} 
  \usepackage{epsfig} 
  \usepackage{graphicx}  
  \usepackage{epstopdf}

 \usepackage[colorlinks=true]{hyperref}

\hypersetup{urlcolor=blue, citecolor=red}
\usepackage{hyperref}

\def\currentvolume{X}
 \def\currentissue{X}
  \def\currentyear{200X}
   \def\currentmonth{XX}
    \def\ppages{X--XX}
     \def\DOI{10.3934/xx.xx.xx.xx}

\usepackage{aimscmds}

\def\cO{{\mathcal{O}}}

 \title[Cell swimming]{The Performance of Discrete  Models of Low Reynolds Number  Swimmers}

 \author[Q.Wang  and H.G. Othmer]{}

 \subjclass{Primary: 92C45, 92C50; Secondary: 92B05}
 
  \keywords{Cell motility, swimming, Low Reynolds number flows}

  \email{qixuanw@uci.edu and othmer@math.umn.edu}

 \thanks{$^{\dagger}$
 $^\ddag$H. G. Othmer is supported in part by NIH Grant \# GM29123-36 and NSF Grant \# 1311974
"Any opinions, findings, and conclusions or recommendations expressed in this
 material are those of the author(s) and do not necessarily reflect the views of
 the National Science Foundation." } 

\begin{document}
\maketitle

\centerline{\scshape Qixuan Wang$\dagger$ }
\medskip
{\footnotesize
   \centerline{Department of Mathematics}
   \centerline{University of California Irvine}
   \centerline{Irvine, CA}
} 

\medskip

\centerline{\scshape  Hans G. Othmer$^{\ddag}$}
\medskip
{\footnotesize
   \centerline{School of Mathematics}
   \centerline{University of Minnesota}
   \centerline{Minneapolis, MN  55445, USA}
} %

\bigskip

\centerline{(Communicated by ???)}

\begin{abstract}
Swimming by shape changes at low Reynolds number  is widely used in biology and
understanding how the efficiency of movement depends on the geometric pattern of
shape changes is important to understand swimming of microorganisms and in designing low 
Reynolds number swimming models. The
simplest models of shape changes are those that comprise a series of linked
spheres that can change their separation and/or their size. Herein we compare
the efficiency of three models in which these modes are used in different ways.

\end{abstract}

\section{Introduction}
\label{intro}

\begin{figure}[htbp]
\centering
\includegraphics[width=.85\textwidth]{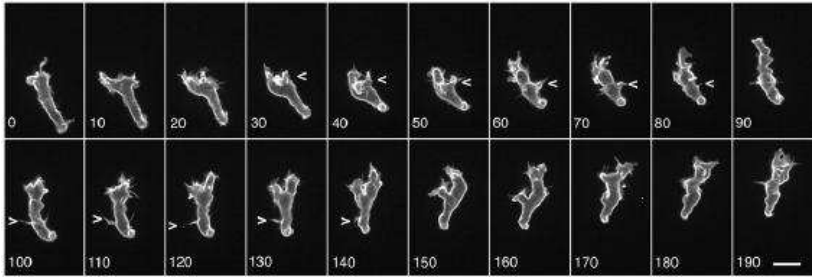}
\caption{A sequence of shape changes observed in a  {\em Dictyostelium
discoideum} cell swimming in a fluid \protect \cite{Barry:2010:DAN}.}
\label{figdd}
\end{figure}

Single-cell organisms use a variety of strategies for translocation, including
crawling, swimming, drifting with the surrounding flow, and others.  Some, such
as bacteria, use flagella, and others, such as paramecia, use cilia to swim, and
both types use only one mode. However other cells can be more flexible in that
they either crawl by transient attachments to their surroundings -- often called
the mesenchymal mode, or by shape changes -- called the amoeboid mode
\cite{Biname:2010:WMC}.  The former may involve strong adhesion to the substrate
or the extracellular matrix (ECM) via integrin-mediated adhesion complexes,
while the latter depends less on force transmission to the ECM or to the  surrounding
fluid, and instead involves shape changes to exploit spaces in the ECM to move
through it (\cf Fig. \ref{figdd}).
The latter mode can be far more effective and can lead to speeds up to forty
times faster than those resulting from mesenchymal motion
\cite{Renkawitz:2010:MFG}.  Cells such as leukocytes, which normally use the
mesenchymal mode in the ECM, can migrate {\em in vivo} in the absence of
integrins, using a 'flowing and squeezing' mechanism \cite{Lammermann:2008:RLM}.
While crawling and swimming are mechanistically distinct strategies, cells can
sense their environment and use the most efficient strategy in a given context.

The spatio-temporal scale of motion of small organisms in viscous fluids
 frequently leads to low Reynolds number (LRN) flows.  Swimmers that use a
 single long, thin flagellum led to the development and application of
 slender-body theory
 \cite{lighthill1976flagellar,hancock1953self,johnson1979flagellar,cox1970motion,batchelor1970slender,keller1976slender,johnson1980improved},
 while microorganisms that swim using a thin layer of cilia were first studied
 by Lighthill for squirming motion of nearly spherical deformable bodies
 \cite{Lighthill:1952:SMN}.  A general review of previous work on swimming
 appears in \cite{Lauga:2009:HSM}, and here we only analyze models that comprise
 a number of linked subunits and have only a finite number of degrees of
 freedom. We call these discrete models of swimmers.

\begin{figure}[htbp]
\centering
\includegraphics[width=1\textwidth]{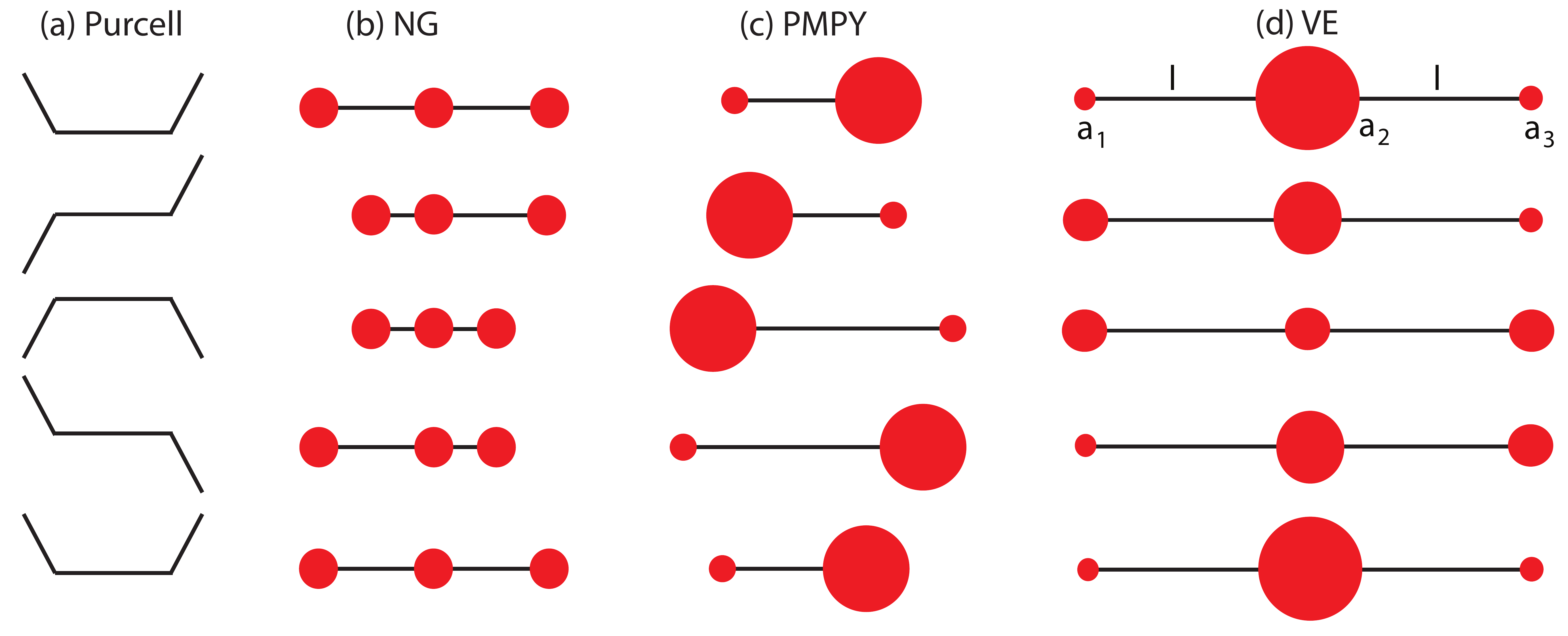}
\caption[Low Reynolds number swimming models]{Low Reynolds number swimming models:
(a) Purcell's 3-link swimmer \cite{purcell1977life};
(b) Najafi-Golestanian's 3-sphere model (NG)\cite{najafi2004simplest};
(c) Pushmepullyou (PMPY) \cite{avron2005pushmepullyou};
(d) The  3-sphere volume-exchange model (VE) \cite{Wang:2012:MLR}.}
\label{fig.intro.4}
\end{figure}

Much of the current interest in locomotion at LRN was stimulated by Purcell's
 description of life at low Reynolds number \cite{Purcell:1977:LLR}. In
 particular, the observation that certain classes of shape changes produce no
 net motion in a viscous fluid led to studies on various types of discrete
 models of swimmers, with the goals of understanding how microorganisms swim and
 facilitating the design of mini-robots that swim at LRN. The first discrete LRN
 model is \textit{Purcell's two-hinge swimmer}, also referred to as
 \textit{Purcell's three-link swimmer} \cite{purcell1977life}
 (Fig.~$\ref{fig.intro.4}$ (a)).  Purcell's model swimmer comprises three
 connected, rigid segments that are constrained to move in a plane and can
 execute restricted rotations around joints linking the segments. The shape is
 specified by two parameters, the angles between adjacent segments, and Purcell
 showed that one can impose sequences of changes in the angles that produce net
 translation of the swimmer. Despite its geometric simplicity, the relationships
 between geometric parameters, speed and efficiency of swimming are not simple
 \cite{becker2003self,avron2008geometric}, but approximations of optimal strokes
 are known \cite{Tam:2007:OSP}.  Various simpler linked-sphere models for which
 both analytical and computational results can be obtained have appeared
 since. The first of these is the Najafi-Golestanian three-sphere  \textit{ accordion}
 model (NG)
 \cite{Najafi:2004:SSL,Golestanian:2007:ART,alexander2009hydrodynamics}
 (Fig.~$\ref{fig.intro.4}$ (b)), which comprises three rigid spheres connected
 by two slender connecting arms aligned along the $x$-direction that can stretch
 and contract in a prescribed form to produce motion.  Since the forces that
 expand or contract the arms are directed along them it can only result in
 translation -- it never rotates. Another linked-sphere model is the
 \textit{pushmepullyou} swimmer (PMPY) \cite{avron2005pushmepullyou}
 (Fig.~$\ref{fig.intro.4}$ (c)), in which two spheres that can expand or
 contract radially are connected by an extensible arm.  Analytical and numerical
 studies of the NG and PMPY models have been done heretofore, and 
 their efficiency and the optimality of various  strokes have been investigated 
  \cite{alouges2008optimal,alouges2009optimal,alouges2011numerical}. Recently we have
 analyzed a three-sphere volume-exchange or  \textit{breather} model (VE) in which the
 spheres are linked by rigid connectors but exchange volume \cite{Wang:2012:MLR}
 (Fig.~$\ref{fig.intro.4}$ (d)), the details of which will be discussed in
 Section~$\ref{linear3sphere}$.

A central problem in the analysis of both biological LRN swimmers and
mini-robots is whether a cyclic sequence of deformations results in significant
movement, and if such a swimming mode is efficient by some measure. This has
been studied for Purcell's swimmer \cite{Tam:2007:OSP}, the PMPY swimmer
\cite{Avron:2004:OSL} and cilia-based swimming
\cite{Michelin:2010:EOS,Osterman:2011:FCB}.  Lighthill's definition
\cite{Lighthill:1952:SMN} provides one  metric of efficiency and several
others have been used, but we will introduce a new criterion to measure the
performance of LRN swimmers.

Of course in reality the various shape changes that have been analyzed require internal
forces that generate the shape changes needed for propulsion of  the organism. In
biological organisms this {\em interior} problem involves the biochemical and
biophysical changes in the cytoskeleton needed to produce the necessary
intracellular forces and shape changes, but here we simply prescribe the shape
changes and treat the {\em exterior} problem.  An integrated model that includes
sensing the environment and controlling the shape changes so as to move is still
beyond reach for even a single-cell organism.

\section{Movement  by shape changes -- the exterior problem} 

  The Navier-Stokes  equations for an incompressible  fluid of density $\rho$, 
 viscosity $\mu$, and velocity $\bfu$ are 
\begin{align}
\label{NSeqn}
\rho\dfrac{\partial \bfu}{\partial t} +\rho (\bfu \cdot\nabla) \bfu &=  \nabla
 \cdot  \stress  + \bff_{\textrm{ext}}  
=- \nabla p  + \mu \Delta \bfu + \bff_{\textrm{ext}} , \\
   \nabla \cdot \bfu &= 0 
\end{align}
where $\stress = -p \bm{\delta} + \mu (\nabla \bfu + (\nabla \bfu)^T)$ is the Cauchy
 stress tensor and ${\bf f}_{\textrm{ext}}  $ is the external force field. 
 Herein we assume that  the swimmer is self-propelled and does not rely on
 any exterior force, and therefore we require
 that $\bff_{\textrm{ext}} = 0$. The Reynolds number based on a characteristic length scale $L$
 and speed scale $U$ is Re = $\rho LU /\mu$, and when converted to
 dimensionless form and the  symbols re-defined, the equations read
\begin{eqnarray}
\label{NSeqn2}
ReSl\dfrac{\partial \bfu}{\partial t} + Re (\bfu \cdot\nabla) \bfu &=& - \nabla p
  +  \Delta \bfu, \nonumber\\   [-7pt] 
&&\\
  \nabla \cdot \bfu &=& 0. \nonumber 
\end{eqnarray}
Here $Sl = \omega L/U $ is the Strouhal number and $\omega$  is a
characteristic frequency of the shape changes. 
When Re$\,\ll1$ the convective momentum  term in (\ref{NSeqn2}) can be
 neglected, but the time variation requires that $ReSl \equiv \omega L^2/\nu \ll 1$.  When both terms
 are neglected, which we assume throughout,  the  flow is governed by the Stokes equations
\begin{equation}
\mu \Delta \bfu  - \nabla p   =  {\bf 0},  \qquad \qquad \nabla \cdot \bfu = 0. 
\label{creep}
\end{equation}
We also only consider the propulsion problem in an infinite domain and impose
the condition $ \mathbf{u}|_{\mathbf{x} \rightarrow \infty} =\mathbf{0}$ on the
velocity.  

In the LRN  regime time does not appear explicitly, momentum is assumed to
 equilibrate instantaneously, and bodies move by exploiting the viscous
 resistance of the fluid. As a result, time-reversible deformations produce no
 motion, which is the content of the `scallop theorem'
 \cite{Purcell:1977:LLR}. In the absence of external forces due to boundaries or
 other fields there is no net force or torque on a self-propelled swimmer in the
 Stokes regime, and therefore movement is a purely geometric process: the net
 displacement of a swimmer during a stroke is independent of the rate at which
 the stroke is executed, as long as the Reynolds number remains small enough.
  Amoebae of the slime mold  \dicty\, have a typical length $L \sim 25
\mu \textrm{m} $ and can swim at $U \sim 3 \mu \textrm{m} / \textrm{min}$
\cite{VanHaastert:2011:ACU}. Assuming the medium is water $(\rho \sim 10^3
\textrm{kg} \ \textrm{m}^{-3}, \ \mu \sim 10^{-3}\, \textrm{Pa} \cdot \textrm{s})$, 
and the deformation frequency $\omega \sim 1 /s$, $Re \sim \cO (10^{-6})$ and
$Sl \sim \cO (10^{-4})$.  In fact the experiments are done in oil that
is significantly more viscous \cite{Barry:2010:DAN}, and for similar cells one
can neglect both inertial terms.  

Suppose that a swimmer occupies the closed compact domain $\Omega (t) \subset
R^n \ (n =2,3) ,$ at time $t$, and let $\partial \Omega (t)$ denote its
prescribed time-dependent boundary. A \textit{swimming stroke} $\Gamma$ is specified by a
time-dependent sequence of the boundary  $\partial \Omega(t)$, and it is
\textit{cyclic} if the initial and final shapes are identical, i.e., $\partial
\Omega (0) = \partial \Omega (T)$ where $T$ is the period
\cite{Shapere:1989:GSP}.  The swimmers' boundary velocity  $\bfV$
relative to fixed coordinates can be written as a part $\bfv$ that defines the
intrinsic shape deformations, and a rigid motion $\bfU$.  If $\mathbf{u}$
denotes the velocity field in the fluid exterior to $\Omega$,  then a standard
LRN self-propulsion problem is : {\em given a cyclic shape deformation specified
by $\bfv$, solve the Stokes equations subject to}
\begin{eqnarray}
\label{eq2}
\int_{\partial \Omega(t)}\stress \cdot \sn = 0, \quad \int_{\partial \Omega(t)}{\mathbf
 {r}}\wedge(\stress \cdot \sn) = 0,   \quad   \mathbf{u} 
 |_{\partial\Omega(t)} = \bfV = \mathbf{v}
+ \bfU, \quad \mathbf{u}|_{\mathbf{x} \rightarrow \infty} =\mathbf{0} 
\end{eqnarray}
where $\sn$ is the exterior normal, and the integrals are the  force- and  torque-free conditions.

In order to treat general shape changes of a cell defined by $\Omega(t) \in R^3$
with boundary $\partial \Omega(t)$, one must solve the exterior Stokes equations
(\ref{creep}) for $\bfu$, with a prescribed velocity $\bfv(t)$ on $\partial \Omega(t)$ and
subject to the decay conditions $\bfu \sim 1/r$ and $p \sim 1/r^{2}$ for $r
\rightarrow \infty$. It is known that the solution has the
representation 
\begin{equation}
\bfu(\bfx) = -\dfrac{1}{8\pi\mu}\int_{\partial \Omega(t)} \bfG(\bfx,\bfy)\cdot\bff(\bfy) dS(\bfy) 
+\dfrac{1}{8\pi}\int_{\partial \Omega(t)} \bfv(\bfy)\cdot\bfT(\bfy,\bfx)\cdot  \sn \,dS(\bfy) 
\label{stsol}
\end{equation}
where $\bfG$ is the free-space Green's function, $\bfT$ is the associated third-rank stress
tensor, and $\bff = \stress \cdot \sn$ is the force on the boundary \cite{Pozrikidis:1992:BIS}.  The
constraints that the total force and the total torque vanish determine the
center-of-mass translational and angular velocities.  When $\bfx \in \partial \Omega(t)$ this
is an integral equation for the force distribution on the boundary, the solution
of which determines the forces needed to produce the prescribed shape changes.
The approach is similar in spirit to what has been done for cells crawling on a
deformable substrate, where the substrate deformations are given and the forces
exerted by the cell are the solution of a Fredholm integral equation
\cite{Barentin:2006:IMC,Butler:2001:TFM}.  

The free space Green's function or Stokeslet has the form 
\begin{equation}
\label{green}
\bfG(\bfx,\bfx_0)  =  \dfrac{1}{r}\left[\bm{I} + \dfrac{\bm{r}\bm{r}}{r^2}\right]
\end{equation}
where $\bm{I}$ is the unit  second-rank tensor, $\bm{r} = \bfx-\bfx_0$, and $r =
|\bfx-\bfx_0|$. Thus the velocity field generated by a point force $\bff$ at the origin is 
\begin{equation}
\label{poinf}
\bm{u}(\bm{x}) = \dfrac{\bfG(\bfx,\bm{0}) }{8 \pi \mu}\cdot \bff
\end{equation}
$\bfG(\bfx,\bm{0}) /(8 \pi\mu)$ is called the Oseen tensor. 
Three other basic solutions that are needed are those for a rigid sphere pulled
through a quiescent fluid, for a radially expanding or contracting sphere, and for the
interaction between two spheres. 

 When a sphere of radius $a$ is pulled through
a quiescent fluid with a steady force $\bfF$ under no-slip conditions at the
surface, the resulting flow field is given by
\begin{equation}
\label{stokesf}
\bm{u}(\bm{r}) = \bfF\cdot (1 + \dfrac{a^2}{6}\nabla^2 )\dfrac{\bfG(\bfx,\bm{x}_s)}{8 \pi
\mu} = \dfrac{\bfF}{8 \pi\mu r}\cdot\left[ \bm{I} + \dfrac{\bm{r}\bm{r}}{r^2} +
\dfrac{a^2}{3r^2}\left[\bm{I} -3\dfrac{\bm{r}\bm{r}}{r^2}\right]\right]
\end{equation} 
where $\bm{x}_s$ is position of the center of the sphere and $\bm{r} = \bfx -
\bfx_s$.  The second term represents the degenerate quadrupole  needed to satisfy the no-slip
boundary condition at $r =a$, but it is small when $a/r \ll 1$. The resulting
velocity of the sphere is given by Stoke's law \cite{Kim:1991:MPS}
\begin{equation}
\label{stokesd}
\bfF = 6\pi\mu a \bm{U}.
\end{equation} 
This can be obtained directly from (\ref{stsol}) by setting $\bfv = 0$ in the
second integral, expanding the Green's function, and defining the total force on
the sphere as 
\begin{equation} 
\bfF = \int_{\partial \Omega} \bff(\bfy) dS(\bfy).
\label{hforce}
\end{equation}

A second basic solution  is  the velocity field $\mathbf{u} $  produced by a
radially expanding sphere, which  can  be generated  by a point source  at the center
$\mathbf{x}_s$ of the sphere \cite{Pozrikidis:1992:BIS}. The corresponding
velocity is 
\begin{equation}
\label{VelocityEqn_7}
\mathbf{u} = \alpha  \dfrac{\mathbf{r}}{r^3}  
\end{equation}
where $\bm{r} = \bfx - \bfx_s$ and $\alpha$ is a constant to be determined. The no-slip boundary condition at
the surface $r =a$  requires that 
\begin{eqnarray}
\label{VelocityEqn_8}
\mathbf{u} (\mathbf{r}) \Big|_{|\mathbf{r}| = a} = \dfrac{\textrm{d} a}{\textrm{d} t} \dfrac{\mathbf{r}}{a}  
\end{eqnarray}
and therefore  $\alpha = \dot{a} a^2$, and 
\begin{eqnarray}\label{VelocityEqn_9}
 \mathbf{u} = \dot{a} \Big( \dfrac{a}{r}  \Big)^2  \dfrac{\mathbf{r}}{r} =
 \dfrac{\dot{v}}{4 \pi r^2} \hat{\mathbf{r}}, 
\end{eqnarray} 
where $v = 4 \pi a^3 /3$ is the volume of the sphere and $ \hat{\mathbf{r}} =
\mathbf{r} / r$.  By combining equations~($\ref{stokesf}$,
$\ref{VelocityEqn_8}$) we obtain the velocity field for the combination of the
pulled and expanding sphere, namely \cite{avron2005pushmepullyou}
\begin{eqnarray}
\label{eq.linear3sphere.1}
\mathbf{u} \big( \mathbf{r}; a, \mathbf{F}, \dot{v} \big) =
\dfrac{1}{24 \pi \mu r} \Big[ \big( 3 + \dfrac{a^2}{r^2} \big) \mathbf{F} +
3 \big( 1 - \dfrac{a^2}{r^2} \big) \big( \mathbf{F} \cdot \widehat{ \mathbf{r} } \big)
\widehat{\mathbf{r}} \Big] + \dfrac{ \dot{v}}{4 \pi r^2} \widehat{\mathbf{r}}.
\end{eqnarray}

The last basic solution needed involves the interaction between two
spheres\footnote{For simplicity we assume that the interactions in a general
configuration of spheres are pairwise additive.}.
Suppose that the $i$th ($i=1,2$) sphere has radius $a_i (t)$, is centered at
$\mathbf{x}_i (t)$ and is subjected to a drag force $\mathbf{F}_i (t)$ due to its
motion. The translational velocity of the $i$th sphere consists of two
parts: $\mathbf{U_{i,0}} = (6 \pi \mu a_i)^{-1} \mathbf{F}_i$ that results
from the drag force $\mathbf{F}_i$ exerted on $i$, and the other a perturbation
part $\delta \mathbf{U}_i$ that is due to the flow generated by the other
sphere. In particular, since the other sphere is translating and expanding,
$\delta \mathbf{U}_i$ can be further decomposed into two parts: $\delta
\mathbf{U}_i^t$ due to the translation of the other sphere, and $\delta
\mathbf{U}_i^e$ that results from its radial expansion. Hence we have the
following decomposition of the translational velocity of the $i$th sphere.
\begin{eqnarray}\label{eq.Int2Sphere_U} 
\mathbf{U}_i = \dfrac{\mathbf{F_i}}{6 \pi \mu a_i} + \delta \mathbf{U}_i^t + \delta \mathbf{U}_i^e
\end{eqnarray}
Here $\delta \mathbf{U}_i^t$ arises   from a flow given by
equation~($\ref{stokesf}$) and is given by 
\begin{eqnarray}  \nonumber
\delta \mathbf{U}_i^t &=& \Big( 1 + \dfrac{a_i^2}{6} \nabla^2 \Big) \mathbf{u} (\mathbf{r})
 \Big|_{\mathbf{r} = \mathbf{x}_i - \mathbf{x}_j } \\   \label{eq.Int2Sphere_deltaUt}
 &=&  \dfrac{1}{8 \pi \mu l} \Big[  \Big( 1 + \dfrac{a_i^2 + a_j^2}{3 l^2} \Big)  \mathbf{I} 
+ \Big( 1 - \dfrac{a_i^2 + a_j^2}{ l^2} \Big) \dfrac{\mathbf{l} \mathbf{l}}{l^2}  \Big] \mathbf{F}_j
\end{eqnarray}
where $\mathbf{l} = \mathbf{x}_i - \mathbf{x}_j$ and $l = |  \mathbf{x}_i -
\mathbf{x}_j |$ \cite{Batchelor:1976:BDP}. The velocity 
$\delta \mathbf{U}_i^e$ is resulted from a flow given by equation~($\ref{VelocityEqn_8}$):
\begin{eqnarray} \label{eq.Int2Sphere_deltaUe}
\delta \mathbf{U}_i^e &=& \Big( 1 + \dfrac{a_i^2}{6} \nabla^2 \Big) \mathbf{u} (\mathbf{r})
 \Big|_{\mathbf{r} = \mathbf{x}_i - \mathbf{x}_j }
 = \mathbf{u} (\mathbf{x}_i - \mathbf{x}_j  ) = \dfrac{a_j^2 \dot{a}_j}{l^3} \mathbf{l}
\end{eqnarray}
Altogether, $\delta \mathbf{U}_i^t$ and $\delta \mathbf{U}_i^3$ induce
a higher-order perturbation in $\bfU$, but as we shall see, these terms are neglected in the
existing analyses of linked-spheres. 

Next we use these solutions in the analysis of various models, and we begin with
the pure volume-exchange (VE) model.

  
\section{The 3-sphere volume-exchange  model}
\label{linear3sphere}

\begin{figure}[h!]
\centering
\includegraphics[height=2.in]{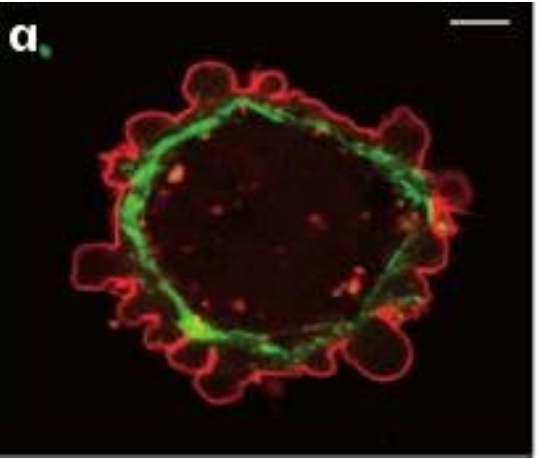}\hspace*{.2in}\includegraphics[height=2.in]{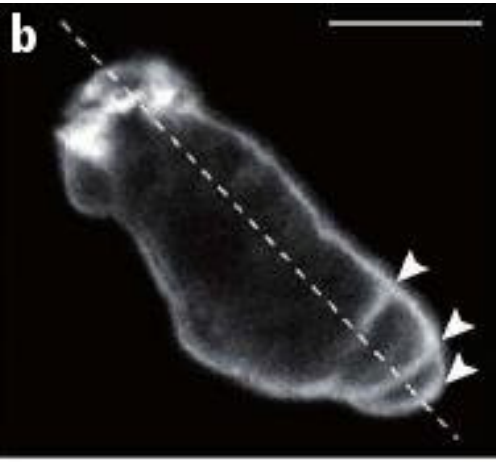} 
\caption{ (a) Blebbing on a melanoma cell: myosin (green) localizes under the
 blebbing membrane (red) (from\protect\cite{Charras:2008:BLW}) (b) The actin
 cortex of a blebbing Dd cell migrating to the lower right. Arrowheads indicate
 the successive blebs and arcs of the actin cortex
 (from\protect\cite{Yoshida:2006:DAM}).}
\label{blebbing}
\end{figure}

Some cells produce membrane protrusions called blebs that emerge when the 
membrane detaches from the cortex locally and the excess internal pressure
forces fluid into the bleb \cite{fackler2008cell,paluch2005cortical}. When this
occurs repeatedly over a cell surface, as in Fig. \ref{blebbing}(a), it may
result in an oscillatory motion of the cell.  In Fig. \ref{blebbing}(a) the cell
blebs blebs profusely with little net translation, whereas Fig. \ref{blebbing}(b) shows a
motile, blebbing {\em Dictyostelium discoideum} cell. If bleb formation is
restricted to the leading edge as in (b), forward motion is driven by
contraction of the cortical network at the rear of the cell. In
either case one can understand the dynamics in terms of mass or volume exchange between
different parts of the cell. The protrusions are usually approximately
hemispherical and thus a linked-sphere model may be a good choice for a study of
blebbing.  However most existing linked-spheres models require significant
changes in the length of the connecting links, which is not 
realistic in blebbing cells. This led us to develop a model that better
describes  blebbing dynamics \cite{Wang:2012:MLR}.

In a minimal VE model there are three spheres linked by two rigid, mass-less
arms of fixed length $l$ (Fig.~$\ref{fig.intro.4}$(d)). A sphere can only expand
or contract in the radial direction (i.e., $a_i = a_i (t)$), and can only
exchange mass with its neighbor(s), {\em i.e.,} 1 with 2 and 2 with 3 but not 1
with 3.  There is little evidence that swimming cells exchange significant
material with the surrounding fluid, and thus we impose mass conservation on the
ensemble. We are primarily concerned with the fluid-structure interaction and
for simplicity we assume that the density in all three spheres is the same
constant, in which case the mass conservation is equivalent to volume
conservation. We also ignore dissipation within cells, and with these
constraints, it is easily seen from the ``scallop theorem''
\cite{Purcell:1977:LLR} that a cyclic two-sphere model cannot swim, since it has
only one degree of freedom.  Therefore a minimal model must comprise at least
three spheres.

The velocity field is given by equation (\ref{eq.linear3sphere.1}), and to obtain
asymptotic solutions, we assume that the distance $l$ between either pair of
neighboring spheres is larger than the radii $a_{i} \ ( i=1,2,3)$, i.e., $a_i  / l \ll 1$, which
simplifies the computation yet captures basic aspects of the movement of
blebbing cells.  The VE model is linear, thus the velocity $\mathbf{U}_i$ of the
$i$th sphere and the force $\mathbf{F}_i$ exerted on the $i$th sphere are all
along the $x$-direction.
\begin{eqnarray*}
\mathbf{U}_i = U_i \mathbf{e}_x, \qquad \mathbf{F}_i = F_i \mathbf{e}_x
\end{eqnarray*}
Because the Stokes problem is linear, the velocity $U_i$ of each sphere is given
by equations~($\ref{eq.Int2Sphere_U}$, $\ref{eq.Int2Sphere_deltaUt}$,
$\ref{eq.Int2Sphere_deltaUe}$), wherein we only retain the leading order terms
in the perturbations $\delta \mathbf{U}_i^t$ and $\delta \mathbf{U}_i^e$.  The
asymptotic solution for the  $U_i$ is given by 
\begin{eqnarray}\label{eq.VE_U1}
U_1 &~\sim& \dfrac{F_1}{6 \pi \mu a_1} + \Big( \dfrac{F_2}{4 \pi \mu l} - \dfrac{\dot{v}_2}{4 \pi l^2} \Big)
+  \Big( \dfrac{F_3}{8 \pi \mu l} - \dfrac{\dot{v}_2}{16 \pi l^2} \Big)
 \\ \label{eq.VE_U2}
U_2 &~\sim& \dfrac{F_2}{6 \pi \mu a_2} + \Big( \dfrac{F_1}{4 \pi \mu l} + \dfrac{\dot{v}_1}{4 \pi l^2} \Big)
+  \Big( \dfrac{F_3}{4 \pi \mu l} - \dfrac{\dot{v}_3}{4 \pi l^2} \Big) \\ \label{eq.VE_U3}
U_3 &~\sim& \dfrac{F_3}{6 \pi \mu a_3} + \Big( \dfrac{F_1}{8 \pi \mu l} + \dfrac{\dot{v}_1}{16 \pi l^2} \Big)
+  \Big( \dfrac{F_2}{4 \pi \mu l} + \dfrac{\dot{v}_2}{4 \pi l^2} \Big)
\end{eqnarray}
where $v_i = 4 \pi a_i^3 / 3$ and $\dot{v}_2 = - \dot{v}_1 - \dot{v}_3$.  Since
the connecting arms have fixed length $l$, $U_1 = U_2 = U_3$, which defines the
translational velocity $U$ of the swimmer.  Since the swimmer is linear it is
necessarily torque-free, and the force-free constraint is
\begin{eqnarray}\label{eq.linear3sphere.19}
 F_1 + F_2 + F_3 = 0.
\end{eqnarray}
The  volume conservation constraint reads 
\begin{eqnarray}\label{eq.linear3sphere.7}
\dot{v}_1 + \dot{v}_2 + \dot{v}_3 = 0 
\end{eqnarray}
and ($\ref{eq.VE_U1} - \ref{eq.linear3sphere.7}$) lead to  the following asymptotic solution for  the 
swimming velocity of the model \cite{Wang:2012:MLR}.
\begin{eqnarray}\label{eq.linear3sphere.10}
U = \dfrac{(a_1 + a_2 - \frac{3}{4} a_3) \dot{v}_1 - (a_3 + a_2 - \frac{3}{4} a_1) \dot{v}_3}{4 \pi l^2 (a_1 + a_2 + a_3)}  
\end{eqnarray}

Next, we consider the power $P \equiv \int_0^T f (t) U(t)\textrm{d} t$ required to propel the swimmer.  The stress on
the surface of the expanding sphere is $\sigma = \mu \dot{v} / (\pi a^3)$
\cite{avron2005pushmepullyou}, and therefore  the power required to expand one sphere is
\begin{eqnarray}
\label{eq.linear3sphere.11}
4 \pi a^2 \sigma \dot{a} = \sigma \dot{v} = \dfrac{4 \mu}{3 v} \dot{v}^2
\end{eqnarray}
Therefore the total instantaneous power expended by the  swimmer in transferring
volumes between the spheres is 
\begin{eqnarray}\label{eq.linear3sphere.12}
P = \dfrac{4 \mu}{3} \Big[ \dfrac{\dot{v}_1^2}{v_1} + \dfrac{\dot{v}_2^2}{v_2} + \dfrac{\dot{v}_3^2}{v_3} \Big] 
= \dfrac{\mu}{\pi} \Big[ \Big( \dfrac{1}{a_1^3} + \dfrac{1}{a_2^3} \Big) \dot{v}_1^2 + \dfrac{2}{a_2^3} \dot{v}_1 \dot{v}_3 + \Big( \dfrac{1}{a_2^3} + \dfrac{1}{a_3^3} \Big) \dot{v}_3^2 \Big].
\end{eqnarray}

Finally we define the performance of a stroke $\Gamma$ as the ratio of the 
 translation per cycle to the  energy expended in a cycle.
\begin{eqnarray}\label{eq.DefPerf}
e = \dfrac{  | \int_0^T  U (t) \textrm{d} t  |}{  \int_0^T P (t) \textrm{d} t} 
\end{eqnarray}
This  has units of 1/force.

When the volume changes are small several conclusions  can be reached  analytically
\cite{Wang:2012:MLR,Wang:2012:MAS}.  Since a swimming stroke is a closed path in
the $v_1 - v_3$ plane or equivalently, a closed path in the $a_1 - a_3$ plane,
we find the following relation between the differential displacement
$\bar{\mathsf{d}} x$ and the differential controls $(\mathsf{d} a_1,
\mathsf{d} a_3)$ from ($\ref{eq.linear3sphere.10}$).
\begin{eqnarray}\label{eq.linear3sphere.14}
\bar{\textrm{d}} X  = \dfrac{\pi}{l^2 } \Big[ a_1^2 \big( 1 - \dfrac{7}{4} \dfrac{a_3}{a_1 + a_2 + a_3} \big) \textrm{d} a_1 - a_3^2 \big( 1 - \dfrac{7}{4}\dfrac{a_1}{a_1 + a_2 + a_3} \big) \textrm{d} a_3 \Big]
\end{eqnarray}
Here $\bar{\textrm{d}} X > 0$ represents an infinitesimal displacement to the
right in Fig.~$\ref{fig.intro.4}(d)$.  The bar in $\bar{\textrm{d}} X$
indicates that the differential displacement is not an exact differential.  

To determine the  direction of swimming, note that from  equation~($\ref{eq.linear3sphere.14})$  we may assume, 
without loss of generality, that $\textrm{d} a_3 = 0$ and $\textrm{d} a_1 > 0$, which means that sphere $3$ does not change, 
and sphere $1$ is expanding while sphere $2$ is contracting. For $a_2$ large enough so that 
$ 1 - 7a_3/(a_1 + a_2 + a_3) > 0$ always holds, we have $\bar{\textrm{d}} X > 0$, 
which means that the swimming direction is from sphere $1$ to sphere $2$.  
Hence we have the following conclusion, which also applies to the PMPY swimmer.
\newtheorem{thm1}{Conclusion}
\begin{thm1}\label{Conc1}
When only one pair of adjacent spheres is involved in volume exchange, and when the 
central  sphere is large enough, the direction of swimming  is always  from the 
expanding sphere to the contracting one.
\end{thm1}

Next, using Stokes' theorem, the translation $\delta X$ corresponding to an infinitesimal closed loop is
\begin{eqnarray}\label{eq.linear3sphere.15}
\delta X = \dfrac{7 \pi}{4 l^2} \Big[ a_1^2 \partial_{a_3} \dfrac{a_3}{a_1 + a_2 + a_3} + a_3^2 \partial_{a_1} \dfrac{a_1}{a_1 + a_2 + a_3}  \Big] \textrm{d} a_1 \wedge \textrm{d} a_3 
\end{eqnarray}
where $\textrm{d} a_1 \wedge \textrm{d} a_3$ denotes the signed area enclosed by the loop. From
this one can show the following.

\newtheorem{thm2}[thm1]{Conclusion}
\begin{thm2}\label{Conc2}
For strokes such that $\Gamma$ is homotopic to the unit circle, increasing
the stroke amplitude will increase the net translation of the stroke, while
increasing the initial radius $a_{20}$ of the central sphere (with $a_{10}$ and
$a_{30}$ unchanged) will decrease the net translation.  Moreover, we have the
approximation
\begin{eqnarray*} 
|X (\Gamma) | \sim \dfrac{\varepsilon }{l} \textrm{Area} (\Omega)
\end{eqnarray*}
where $\varepsilon \sim a_i/l$ and $\Omega$ is the region enclosed by $\Gamma$ and $\textrm{Area} (\Omega)$ is its signed area.
\end{thm2}
The proof  of this  is given in \cite{Wang:2012:MAS}.

 \subsection{Numerical computations} Next we prescribe cycles of shape changes
in the controls $(\dot{a}_1, \dot{a}_3)$ and compute the displacement and
performance measure numerically.  In particular, we investigate how the
following characteristics of the system affect the net translation $X = \int_0^T
U(t) \textrm{d} t$ and the performance $e$ of the swimmer
after a full cycle ($T =1$).
\begin{enumerate} 
 \item $L$, which measures the distance between a pair of neighboring sphere;
 \item  $r_1$, $r_2$, which measure the amplitude of shape deformations;
 \item  $s$, which measures the size of the central sphere.
 \end{enumerate}

 We use the following protocol in varying these parameters. 
 \begin{eqnarray*}
 \textrm{Arm length (fixed) :} & & l_1 = l_2 =L \\
 \textrm{Controls:}  & &  a_1 (t) = R_0 + r_1 \cos 2 \pi t, \quad a_3 (t) =  R_0 + r_3 \sin 2 \pi t, \\
 \textrm{Radius of the central sphere:} & & a_2 (0) = s, \quad a_2 (t) = \big( \dfrac{3}{4 \pi} V_{\textrm{tot}} - a_1 (t)^3 - a_3 (t)^3  \big)^{\frac{1}{3}} 
 \end{eqnarray*}

We  consider three cases.

   \begin{figure}[htbp]
\centering
\includegraphics[width=1\textwidth]{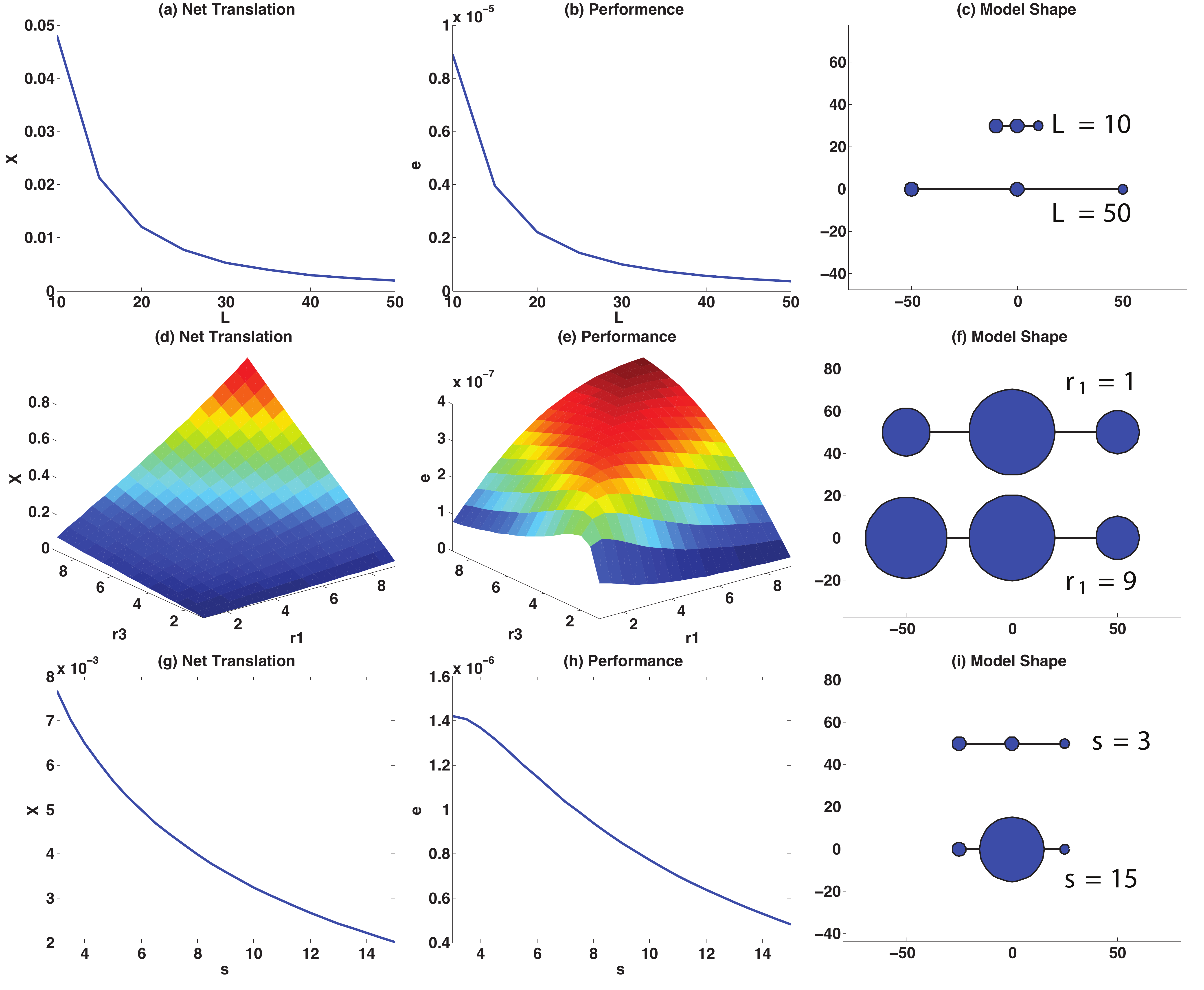}
\caption{(a,b) Translation $X$ and performance $e$ as a function of the fixed arm length $L$. (c) The initial
shape of the swimmer when $L = 10$ or $L = 50$ in simulations shown in a\& b.
(d,e) Translation $X$ and performance $e$ as a function  of the stroke amplitude $(r_1, r_3)$. (f) The initial shape 
of the swimmer when $r_1 = 1$ or $r_1 = 9$ in simulations shown in (d,e). Notice that although both $r_1$ and
$r_3$ determine the stroke amplitude, the initial shape is only related to $r_1$. (g,h) Translation $X$ and 
performance $e$ in relation of the (initial) size of the central sphere $s$. (i) The initial shape 
of the swimmer when $s=3$ or $s=15$ in simulations shown in (g,h). The scales are the same for panels
(c,f,i).}
 \label{fig.VE3Sphere_Lrs}
\end{figure}
 
 \begin{itemize}
\item [{\bf I}]  \normalfont First, we set $r_1 = r_3 = 1$, $R_0 = 2$ and $s=3$, and test different values
 of $L$, i.e., arm length (Fig.~$\ref{fig.VE3Sphere_Lrs}$(a,b)). Fig.~$\ref{fig.VE3Sphere_Lrs}$(c) gives the initial profile
 of the swimmers with $L=10$ and $L=50$, respectively. Fig.~$\ref{fig.VE3Sphere_Lrs}$(a,b) show that both the translation $X$ and
 the performance $e$ decrease as the arm length increases. That is to say, with a fixed amount of 
 body mass, a too long body is not a good strategy for swimming. The effect on the translation $X$ 
 can be seen from equation~($\ref{eq.linear3sphere.10}$)
 and Conclusion~$\ref{Conc2}$, with a fixed stroke rending the same Area$(\Omega)$, the translation $X$ scales 
 in the order of $\varepsilon l^{-1} = a l^{-2}$. Hence as the arm length $L$ increases, $X$ decreases quickly.
As for performance $e$ (equation~($\ref{eq.DefPerf}$)), it is the ratio of translation to energy over a cycle, with 
power $P$ given by equation~(\ref{eq.linear3sphere.12}), from which we clearly see that the arm length $L$ does
not enter into the expression of $P$. Hence $e$ decays similarly to $X$, namely, in the order of $ l^{-2}$.

\item [{\bf II}] Next, we set  $L = 50, s=10, R_0 = 10$, and test different values of $r_1$ and
$r_3$ (Fig.~$\ref{fig.VE3Sphere_Lrs}$(c,d)), i.e., the stroke amplitude. Fig.~$\ref{fig.VE3Sphere_Lrs}$(e) gives the initial profile
 of the swimmers with $r_3 = 1$ always, but $r_1 =1$ or $r_1 = 9$, respectively.
 From Fig.~$\ref{fig.VE3Sphere_Lrs}$(c) we see that
  The translation increases as
either $r_i$ increases, but if one of $r_i$ is small, the increase of
translation due to the other $r_j$ is small. This can be explained by Conclusion~\ref{Conc2}, in general a large $r_1$
or/and $r_3$ indicate a large stroke, i.e., a large Area$(\Omega)$, which clearly induces a large translation $X$ ---
except that when one of $r_i$ is small, then no matter how large the other $r_i$ is, we have Area$(\Omega) \sim 0$, which
results in a $X \sim 0$. As for performance, it is difficult to obtain a general analysis between $e$ and Area$(\Omega)$, yet
Fig.~$\ref{fig.VE3Sphere_Lrs}$(c) indicates that $e$ has the similar behavior as $X$ (though maybe not to the 
same order), i.e., $e$ increases as either $r_i$ increases, if the other $r_i$ is not too small.

\item [{\bf III}] Finally we set  $L = 25, r_1 = r_3 = 1$ and $R_0 = 2$, and test different
 values of $s$ (Fig.~$\ref{fig.VE3Sphere_Lrs}$(e,f)), i.e., the (relative) size of the central sphere to the side spheres.
Here we would like to point out such a test can be considered a complement to the test of stroke amplitude. Consider that
we may scale the total volume of the whole object to be the same with respect to different values of $s$, then a model
with a small central sphere (Fig.~$\ref{fig.VE3Sphere_Lrs}$(i), $s = 3$) is translated to one with a large stroke, and on the 
other hand a large central sphere (Fig.~$\ref{fig.VE3Sphere_Lrs}$(i), $s = 15$) corresponds to a swimmer with a small
stroke. Hence the results can be predicted from previous discussion and are shown in Fig.~$\ref{fig.VE3Sphere_Lrs}$(e,f):
both the translation and the performance decrease as the initial radius of the central sphere is increased. Yet we illustrate 
such behavior in regard to $s$. Observe equation~($\ref{eq.linear3sphere.15}$) and we find that
with both the stroke amplitude $\textrm{d} a_1 \wedge \textrm{d} a_3 $ and the arm length $L$
fixed, $X$ is proportional to the following quantity:
\begin{eqnarray}\label{eq.AnalysisXs}
a_1^2 \partial_{a_3} \dfrac{a_3}{a_1 + a_2 + a_3} + a_3^2 \partial_{a_1} \dfrac{a_1}{a_1 + a_2 + a_3}
\end{eqnarray}
which clearly decreases as the central sphere gets bigger. On the other hand, equation~($\ref{eq.linear3sphere.12}$) indicates that
a bigger central sphere results in small power $P$. To evaluate the effect of the size of the central sphere on the performance $e$,
we conduct an asymptotic analysis with the scenario $a_1, a_3 \ll a_2$. In such a case, the quantity given by equation~($\ref{eq.AnalysisXs}$)
approximates $0$, then equation~($\ref{eq.linear3sphere.15}$) shows that $\delta X \sim 0$. However from equation~($\ref{eq.linear3sphere.12}$)
we obtain the following behavior $P$ when $a_1, a_3 \ll a_2$:
\begin{eqnarray*}
P \sim \dfrac{\mu}{\pi} \Big[  \dfrac{1}{a_1^3} \dot{v}_1^2 +  \dfrac{1}{a_3^3} \dot{v}_3^2  \Big]  
\end{eqnarray*}
which in general does not vanish. Hence as the ratio of average translation to average power, the performance $e$ vanishes in the scenario $a_1, a_3 \ll a_2$.

 \end{itemize}

 In conclusion, in order to reach longer net translation $X$ or better
 performance $e$, the 3-sphere volume-exchange model should be designed so that
 \textit{the connecting arm is short, more mass are exchanged among the spheres
 though the cycle, and the central sphere should not be big comparing to the two
 side spheres.}

 \section{A comparison of the three linked-sphere swimmers}
 
 Here we compare the performance of the three linked-sphere swimmers: the  NG 3-sphere model, the pushmepullyou
 (PMPY) and the VE model. A summary of the analysis that leads to the  velocity and power of 
 the NG and PMPY models can be found in Appendix $\ref{Sec.ReviewModel}$ and in 
 \cite{najafi2004simplest,golestanian2007analytic,avron2005pushmepullyou}.  To
 standardize the  comparison between  them, we stipulate  that the total volume in all
 spheres are the same for each model, and the stroke amplitudes are the same. We
 prescribe strokes for each model as follows. 
\begin{itemize}
\item \textbf{NG:}
 \begin{eqnarray*}
 & &R_1 = R_2 = R_3 = 2 r_{\textrm{G}} \\
 & & l_1 (t) = L + \cos (2 \pi t), \quad l_2 (t) = L + \sin (2 \pi t)
 \end{eqnarray*}
 \item \textbf{Pushmepullyou:} 
\begin{eqnarray*}
 & &R_1 = 2 r_{\textrm{P}}  + \cos (2 \pi t) \\
 & & R_2 (0) = 2r_{\textrm{P}}, \quad R_1^3 + R_2^3 = \dfrac{3}{4 \pi} V_{\textrm{tot}}  \\
 & & l (t) = L +   \sin (2 \pi t)
 \end{eqnarray*}
 \item \textbf{Volume-exchange (VE):} 
\begin{eqnarray*}
 & & l_1 = l_2 = L \\
 & & R_1 (t) = 2 r_{\textrm{V}}+ \cos (2 \pi t) , \quad R_3 (t) = 2 r_{\textrm{VE}} + \sin (2 \pi t) \\
 & &R_2(0) = 2 r_{\textrm{V}}, \quad R_1^3 + R_2^3 + R_3^3 = \dfrac{3}{4 \pi} V_{\textrm{tot}} 
 \end{eqnarray*}
 \end{itemize}
 where scales $r_{\textrm{G}} , r_{\textrm{P}} ,r_{\textrm{V}}$ are chosen so that the total volume
 of all spheres in each swimmer are the same. Without loss of generality, 
 we may choose $ r_{\textrm{G}}  = 1$.

\begin{figure}[htbp]
\centering
\includegraphics[width=1\textwidth]{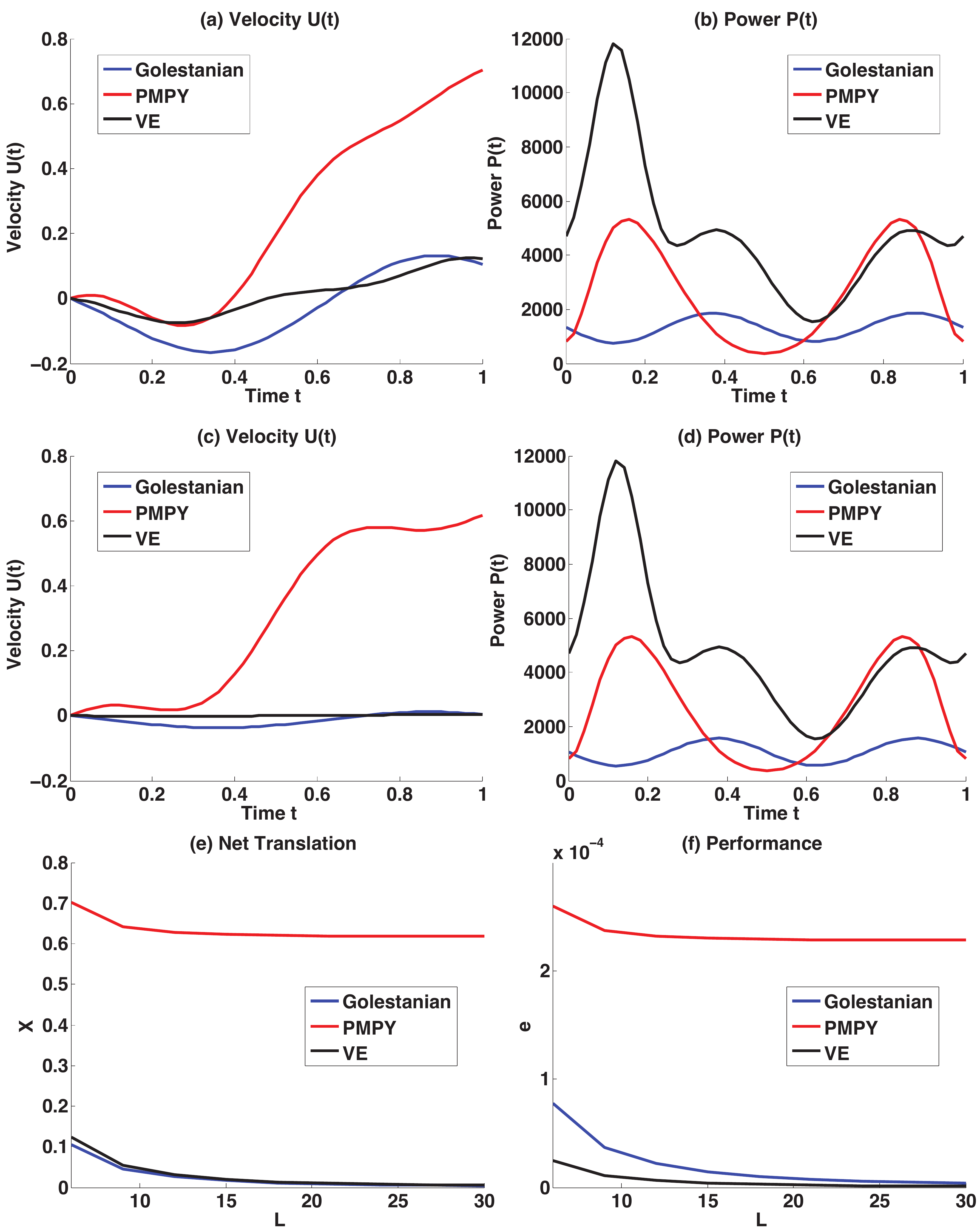}
\caption{A comparison of  the three linked-sphere swimmers.
(a,b) $U(t)$ and $P(t)$ of the swimmers within a cycle with $L = 6$.
(c,d) $U(t)$ and $P(t)$ of the swimmers within a cycle with $L = 30$.
(e,f)  Net translation $X$ and performance $e$ of the 
three swimmers with different values of $L$.}
 \label{fig.Comp3Swimmer}
\end{figure}

 First, the velocity  $U (t)$ and the power $P (t)$ within a cycle that result from the above prescribed
 strokes are given in Fig.s~$\ref{fig.Comp3Swimmer}(a-d)$, for $L = 6 $ or
 $30$, respectively.  For translation, comparing
 Fig.~$\ref{fig.Comp3Swimmer}$(a) and (c) we see that $U (t)$ for PMPY does
 not change much, while it almost vanishes for NG  and VE when $L=30$.
 On the other hand, $P (t)$ in Fig.~$\ref{fig.Comp3Swimmer}$(b) and
 Fig.~$\ref{fig.Comp3Swimmer}$(d) are quite much similar.

Moreover, we find that the PMPY model is clearly superior to the others
 (Fig.~$\ref{fig.Comp3Swimmer}$(e,f)), in regard to both translation and
 performance over the entire range of L, even when $L\rightarrow \infty $.  The
 reason can be found in the asymptotic solutions of $U$ and $P$ for the three
 swimmers. While the leading term of $U$ for PMPY is ${\mathcal{O}} (1)$, it is
 ${\mathcal{O}} (l^{-1})$ and ${\mathcal{O}} (l^{-2})$ for NG and VE,
 respectively.  The leading order term in the power $P$ is $\cO (1)$ (equation~($\ref{eq.linear3sphere.12}$)) for all three
 swimmers, hence the arm length $L$ does not have much influence on the power or 
 the performance $e$. The
 leading order term of $e$ is the same as that of $U$, i.e., $\cO(1),
 \cO(l^{-1}), \cO(l^{-2})$ for PMPY, Golestanian and VE, respectively.

\section{Mixed controls result in more net translation and better performance}

All three models  have two degrees of freedom, which are of two types -- a change
in the arm length $\dot{l}$, or a change in the sphere radius
$\dot{a}$. Different combinations of the controls result in different swimming
behaviors, and from the results above we find that the best choice is the mixed
strategy, {\em i.e.}, $(\dot{l}, \dot{R})$, which is adopted by PMPY. A
combination of the same kind --- $(\dot{l}_1, \dot{l}_2)$ or $(\dot{R}_1,
\dot{R}_2)$ --- is not advantageous, and they produce comparable net translation
and performance (Fig.~$\ref{fig.Comp3Swimmer}$).

The priority of the mixed strategy is based on the following two principles:
\begin{enumerate}
 \item $\dot{l}$ results in a velocity with leading order $\cO(1)$,  while the
       leading order that results from $\dot{R}$ is only $\cO 
(l^{-2})$. Hence to increase the  net translation, one should incorporate a
change in arm length.

\item The leading order terms in the velocity $U (t)$ should not be, or even
      approximate, an exact differential, because  they then make no or little contribution to the
net translation.  
\end{enumerate}

We have discussed the first principle above, which clearly explains why PMPY is
better than the other two swimmers. However this does not explain why the net
translation behaves similar for NG and VE, with leading order terms that scale like
$\cO (l^{-1})$ and $\cO (l^{-2})$, respectively. In fact, it is only in the case that all
three spheres are of equal size that  the velocity scales like $\cO (l^{-1})$ for the 
NG swimmer. When tspheres are of different sizes are involved, $U (t)$ will
actually scale as $\cO (1)$, {\em i.e.}, the same as $U (t)$ in the PMPY model. However,
  the net translation $X$ still turns out to be of order $\cO (l^{-2})$,
which is the same as for the VE swimmer.

To understand this, we analyze the  asymptotic solution of $U (t)$ for an 
NG swimmer which is approximated to the order $\cO (l^{-1})$ (equation~(A1),
\cite{golestanian2007analytic}). The equation is complex, but it can be written
in the following form.
\begin{eqnarray}\label{eq.UG_Algebra}
U = A_0 \dot{l}_1 + B_0 \dot{l}_2 + \Big( \dfrac{A_1^{(1)}}{l_1}  + \dfrac{A_1^{(2)}}{l_2} + \dfrac{A_1^{(12)}}{l_1 + l_2}  \Big)  \dot{l}_1 
+ \Big( \dfrac{B_1^{(1)}}{l_1}  + \dfrac{B_1^{(2)}}{l_2} + \dfrac{B_1^{(12)}}{l_1 + l_2}  \Big)  \dot{l}_2 + \cO (\dfrac{1}{l^2})
\end{eqnarray}
where all coefficients $A_i^{(\alpha)}, B_i^{(\alpha)}$ are functions of $a_1,
a_2, a_3$ only and do not depend on $l_1, l_2$ or time $t$. 

The leading order term of $U$, denoted as $U_{(0)}$, is the combination
\begin{eqnarray*}
U_{(0)}= A_0 \dot{l}_1 + B_0 \dot{l}_2
\end{eqnarray*}
and the integral over a whole cycle gives
\begin{eqnarray}
\label{eq.Golestanian_X0}
X_{(0)}= \int_0^T U_{(0)} \ \textrm{d} t = \int_0^T \big(A_0 \dot{l}_1 + B_0 \dot{l}_2 \big) \ \textrm{d} t 
= A_0 \dot{l}_1 \big|_{t = 0}^{t = T} + B_0 \dot{l}_2 \big|_{t = 0}^{t = T} = 0.
\end{eqnarray}
Next, the $\cO (l^{-1})$ term of $U$, which we denote $U_{(1)}$, is given by the
following. 
\begin{eqnarray*}
U_{(1)} = \Big( \dfrac{A_1^{(1)}}{l_1}  + \dfrac{A_1^{(2)}}{l_2} + \dfrac{A_1^{(12)}}{l_1 + l_2}  \Big)  \dot{l}_1
+ \Big( \dfrac{B_1^{(1)}}{l_1}  + \dfrac{B_1^{(2)}}{l_2} + \dfrac{B_1^{(12)}}{l_1 + l_2}  \Big)  \dot{l}_2
\end{eqnarray*}
In general, the integral $\int_0^T U_1 \ \textrm{d} t$ does not vanish, but  we
have the relation
\begin{eqnarray*}
l_i (t) = L + \delta l_i (t)
\end{eqnarray*}
where $L$ is the fixed part and $\delta l_i$ is the deformation part. When  $L $
is sufficiently large, so as to ensure that the higher-order interactions
between spheres are negligible,  we have that  $\delta l_i \ll  L$, and thus
$U_{(1)}$ can be approximated as
\begin{eqnarray*}
U_{(1)} \sim \Big( \dfrac{A_1^{(1)}}{L}  + \dfrac{A_1^{(2)}}{L} + \dfrac{A_1^{(12)}}{2 L}  \Big)  \dot{l}_1
+ \Big( \dfrac{B_1^{(1)}}{L}  + \dfrac{B_1^{(2)}}{L} + \dfrac{B_1^{(12)}}{2 L}  \Big)  \dot{l}_2
\end{eqnarray*}
Thus again,
\begin{eqnarray}
\label{eq.Golestanian_X1} 
\nonumber
X_{(1)} &\sim& \int_0^T U_{(1)} \ \textrm{d} t  \\   \nonumber
&=&  \Big( \dfrac{A_1^{(1)}}{L}  + \dfrac{A_1^{(2)}}{L} + \dfrac{A_1^{(12)}}{2 L}  \Big) \int_0^T \dot{l}_1   \ \textrm{d} t 
+ \Big( \dfrac{B_1^{(1)}}{L}  + \dfrac{B_1^{(2)}}{L} + \dfrac{B_1^{(12)}}{2 L}  \Big)  \int_0^T \dot{l}_2  \ \textrm{d} t  \\   
&=& 0
\end{eqnarray}

From equations~($\ref{eq.Golestanian_X0}, \ref{eq.Golestanian_X1}$), we see that
although in general the $\cO(1)$ and $\cO (l^{-1})$ terms do not vanish in $U (t)$, 
the $\cO (1)$ term is an exact differential and  the $\cO (l^{-1})$ term is
approximately an exact differential, and hence the net translation $X \sim \cO
(l^{-2})$. This 
is the same as the leading order that results from varying the radii in the VE model.

\begin{figure}[htbp]
\centering
\includegraphics[width=1\textwidth]{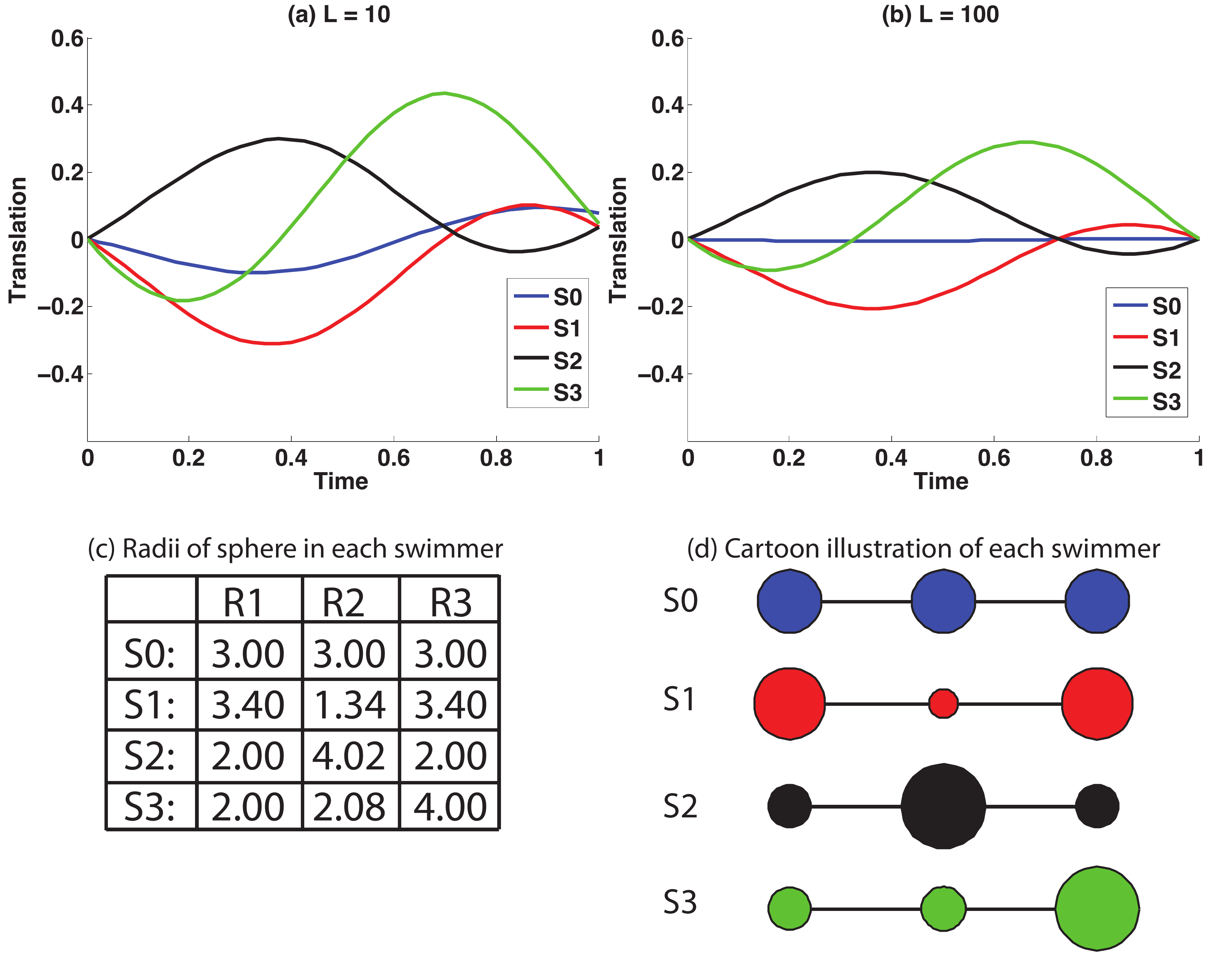}
\caption{(a) $U(t)$ of four Golestanian swimmers, with $L=10$. 
(b) $U(t)$ of four Golestanian swimmers, whose spheres have correspondingly the same size as in (a),but with
$L=10$. (c) The radius of each sphere in each swimmer of (a,b).
(d) Initial shapes of the four swimmer (with $L = 15$ for better visualization
effects).  }
 \label{fig.Golestanian_U_L}
\end{figure}

Fig.~$\ref{fig.Golestanian_U_L}$(a,b) give $U (t)$ within one cycle for four
NG swimmers, whose radii are given in Fig.~$\ref{fig.Golestanian_U_L}$(c),
with $L=10$ or $L=100$. Again, to make a fair comparison we require the total
volume of all three spheres in each swimmer are the same. We do not use
equation~(A1) in \cite{golestanian2007analytic} to solve for $U (t)$, instead
we numerically solve the whole system (equations~($\ref{eq.NG_U1}$ - $\ref{eq.NG_U3}$)).  From
Fig.~$\ref{fig.Golestanian_U_L}$(a,b) we see that for the same swimmer, the
amplitude of $U (t)$ within a cycle is almost of the same scale when $L = 10$ or
$100$, yet the net translation is very small with either value of $L$. However
we do observe that among different choices of the sphere sizes, net translation
favors the equal sized spheres (S0) -- {\em i.e.}, S0 results in the most net
translation yet it requires the least amplitude of $U(t)$ among the four
swimmers.

On the other hand, for PMPY the  leading term in $U (t)$ is 
\begin{eqnarray*} 
U_{(0)} = \dfrac{a_1 - a_2}{2 ( a_1 + a_2)} \dot{l} \sim \cO (1)
\end{eqnarray*}
which is not an exact differential, and as a result, the leading order term
in the net translation gives
\begin{eqnarray*} 
X_{(0)} = \int_0^T \dfrac{a_1 - a_2}{2 ( a_1 + a_2)} \dot{l} \  \textrm{d} t  \sim \cO (1)
\end{eqnarray*}
in general, which explains the better performance of the PMPY swimmer 
as compared with the  Golestanian or VE swimmer.

\section{Discussion} 

We have compared the three most widely-studied discrete swimmers and have shown
that the PMPY swimmer performs best under the imposed conditions. This
conclusion is important for the design of mini-robots, but of course real
systems are more complex, and as we indicated earlier, the VE model is a more
realistic model of cellular motion. An analysis of continuum models of swimmers
that more accurately reflect actual shape changes will be reported elsewhere.

\appendix
\section{A summary  of the NG and PMPY models}
\label{Sec.ReviewModel}

The NG swimmer (Fig.~$\ref{fig.intro.4}$(b)) consists of three spheres with
radii $a_i \ (i=1,2,3)$ and two connecting arms with length $l_i (t)\ (i=1,2)$
\cite{Najafi:2004:SSL,Golestanian:2007:ART,alexander2009hydrodynamics}.  While
the spheres are rigid, the connecting arms can stretch or contract. In the case
when $a_i / l \ll 1$, the velocities of the spheres ($U_i$) are related to the
forces exerted on the spheres ($F_i$) via the Oseen tensor:
\begin{eqnarray}\label{eq.NG_U1}
U_1 &=& \dfrac{F_1}{6 \pi \mu a_1} + \dfrac{F_2}{4 \pi \mu l_1} + \dfrac{F_3}{4 \pi \mu (l_1 + l_2)}  \\ \label{eq.NG_U2}
U_2 &=& \dfrac{F_1}{4 \pi \mu l_1} + \dfrac{F_2}{6 \pi \mu a_2} + \dfrac{F_3}{4 \pi \mu l_2}  \\ \label{eq.NG_U3}
U_3 &=& \dfrac{F_1}{4 \pi \mu (l_1 + l_2)} + \dfrac{F_2}{4 \pi \mu l_2} + \dfrac{F_3}{6 \pi \mu a_3}
\end{eqnarray}
The velocities are related via the following relations:
\begin{eqnarray}\label{eq.NG_URelation}
U_2 - U_1 = \dot{l}_1, \qquad U_3 - U_2 = \dot{l}_2
\end{eqnarray}
The system is force-free
\begin{eqnarray}\label{eq.NG_ForceFree}
F_1 + F_2 + F_3 = 0
\end{eqnarray}
and equations~($\ref{eq.NG_U1}$ - $\ref{eq.NG_ForceFree}$) are  a closed system. The velocity of the 
whole swimmer is defined as
\begin{eqnarray}\label{eq.NG_U_Def}
U = \dfrac{1}{3} (U_1 + U_2 + U_3)
\end{eqnarray}
In the case that the spheres are of equal size, we have the asymptotic solution
\begin{eqnarray}\label{eq.NG_U_EqSize}
U = \dfrac{a}{6} \Big[  \dfrac{\dot{l}_2 - \dot{l}_1}{l_1 + l_2} 
+ 2 \Big( \dfrac{\dot{l}_1}{l_2} -  \dfrac{\dot{l}_2}{l_1} \Big)  \Big].
\end{eqnarray}
The power consumption of the swimmer comes from dragging the spheres, thus
\begin{eqnarray}\label{eq.NG_P_Def}
P = F_1 U_1 + F_2 U_2 + F_3 U_3  
\end{eqnarray}
and in the case that the spheres are of equal size the above equation simplifies to
\begin{eqnarray}\label{eq.NG_P_EqSize}
\dfrac{P}{4 \pi \mu a} &=& \Big[ 1 + \dfrac{a}{l_1} - \dfrac{a}{2 l_2} + \dfrac{a}{l_1 + l_2} \Big] \dot{l}_1^2  
  + \Big[ 1 - \dfrac{a}{2 l_1} + \dfrac{a}{l_2} + \dfrac{a}{l_1 + l_2} \Big] \dot{l}_2^2 \\
& & + \Big[ 1 - \dfrac{a}{2 l_1} - \dfrac{a}{2 l_2} + \dfrac{5 a}{2 (l_1 + l_2)} \Big] \dot{l}_1 \dot{l}_2.
\end{eqnarray}

The PMPY swimmer (Fig.~$\ref{fig.intro.4}$(c)) consists of two spheres with
radii $a_i (t) \ (i=1,2)$ and one connecting arms with length $l (t) $.  The
spheres can expand or contract in the radial direction, and the connecting arms
can stretch or contract. When $a_i / l \ll 1$, the velocities of the
spheres ($U_i$) are related to the forces exerted on the spheres ($F_i$) via the
Oseen tensor
\begin{eqnarray}\label{eq.PMPY_U1}
U_1 &=& \dfrac{F_1}{6 \pi \mu a_1} + \dfrac{a_1^2 }{l^2} \dot{a}_1  \\   \label{eq.PMPY_U2}
U_2 &=& \dfrac{F_2}{6 \pi \mu a_2} + \dfrac{a_2^2 }{l^2} \dot{a}_2.
\end{eqnarray}
The velocities are related via the following relation.
\begin{eqnarray}\label{eq.PMPY_URelation}
U_2 - U_1 = \dot{l} 
\end{eqnarray}
Again, the system is force-free
\begin{eqnarray}\label{eq.PMPY_ForceFree}
F_1 + F_2  = 0
\end{eqnarray}
and the total volume of the two spheres is conserved
\begin{eqnarray}\label{eq.PMPY_Volume}
a_1^2 \dot{a}_1 + a_2^2 \dot{a}_2 = 0.
\end{eqnarray}
Equations~($\ref{eq.PMPY_U1}$ - $\ref{eq.PMPY_Volume}$) are a closed system and the velocity of the 
swimmer is
\begin{eqnarray}\label{eq.PMPY_U}
U = \dfrac{1}{2} (U_1 + U_2 ) = \dfrac{a_1 - a_2}{2 (a_1 + a_2)} \dot{l} + \dfrac{a_1^2 }{l^2} \dot{a}_1. 
\end{eqnarray}
The power consumption $P (t)$ of the swimmer comprises two parts: $P_{\textrm{drag}}$ that results from
the drag force on the spheres, which is given by
\begin{eqnarray}\label{eq.PMPY_PowerDrag}
P_{\textrm{drag}} = F_1 U_1 + F_2 U_2
\end{eqnarray}
and $P_{\textrm{exp}}$ that results from the radial expansion of the swimmers
\begin{eqnarray}\label{eq.PMPY_PowerExp}
P_{\textrm{exp}} = 16 \pi \mu (a_1 \dot{a}_1^2 + a_2 \dot{a}_2^2).
\end{eqnarray}
Hence the power expended is given by 
\begin{eqnarray}\label{eq.PMPY_Power}
P = 6 \pi \mu \Big( \dfrac{1}{a_1} + \dfrac{1}{a_2} \Big)^{-1} \dot{l}^2 + 16
\pi \mu (a_1 \dot{a}_1^2 + a_2 \dot{a}_2^2).
\end{eqnarray}

\flushleft{\bf References}
 
\bibliographystyle{xauthordate1}
\bibliography{a-e,f-l,m-r,s-z,newrefs,thesis}    

\medskip


Received\\
Accepted 

\end{document}